\documentclass[twocolumn,10pt]{confpaper}
\usepackage{usenix,epsfig,url,amssymb,CJKutf8}
\usepackage{balance}
\usepackage{titling}
\begin{document}

%make title bold and 14 pt font (Latex default is non-bold, 16 pt)

\title{\fontsize{14}{14} \textbf{Automatically Generating a Large, \\ Culture-Specific Blocklist for China}}
\date{}
%for single author (just remove % characters)
\author{
{\rm Austin Hounsel}\\
Princeton University
\and
{\rm Prateek Mittal}\\
Princeton University
\and
{\rm Nick Feamster}\\
Princeton University
% copy the following lines to add more authors
% \and
% {\rm Name}\\
%Name Institution
} % End author

\posttitle{\par\end{center}}
\setlength{\droptitle}{-10pt}
\maketitle

% Use the following at camera-ready time to suppress page numbers.
% Comment it out when you first submit the paper for review.
% \thispagestyle{empty}
% \pagestyle{empty}

\begin{abstract}
Internet censorship measurements rely on lists of websites to be
tested, or ``block lists'' that are curated by third parties.
Unfortunately, many of these lists are not public, and those that are
tend to focus on a small group of topics, leaving other types of sites
and services untested. To increase and diversify the set of sites on
existing block lists, we use natural language processing and search
engines to automatically discover a much wider range of websites that
are censored in China. Using these techniques, we create a list of 1125
websites outside the Alexa Top 1,000 that cover Chinese politics,
minority human rights organizations, oppressed
religions, and more. Importantly, \textit{none of the sites we discover are
present on the current largest block list}. The list that we develop
not only vastly expands the set of sites that current Internet
measurement tools can test, but it also deepens our understanding of
the nature of content that is censored in China. We have released both
this new block list and the code for generating it.
\end{abstract}

\section{Introduction} 

Internet censorship is pervasive in China. Topics ranging from
political dissent and religious assembly to privacy-enhancing
technologies are known to be
censored~\cite{citizenlab:block}. However, the Chinese government has
not released a complete list of websites that they have
filtered.~\cite{fhouse:china}. When performing measurements of
Internet filtering, then, the inability to know what sites are blocked
creates a circular problem of discovering the sites to measure in the
first place. To do so, various third parties currently curate lists of
websites that are known to be censored, or ``block lists''. These
lists are used to both understand {what} content is censored in China
and how that censorship is implemented. Indeed, the Open Observatory
of Network Interference notes that ``censorship findings are only as
interesting as the sites and services that you
test.''~\cite{ooni:lists}.

Instead of curating a block list by hand, Darer et al. proposed a
system called FilteredWeb that automatically discovers web pages that
are censored in China~\cite{darer2017filteredweb}. Their approach is
summarized in the following steps. First, keywords are extracted from
web pages on the Citizen Lab block list, a small, hand-curated
list. These are English words that are ranked through TF-IDF, a
technique which we describe in Section ~\ref{tf-idf}. Then, each
keyword is used as a query for a search engine, such as Bing. The
intuition is that censored web pages contain similar
keywords. Finally, each web page that appears in the search results is
tested for DNS manipulation in China. These tests are performed by
sending DNS queries to IP addresses in China that don't belong to DNS
servers. If a DNS response is received, then, it is inferred that the
request was intercepted in China and that the website is
censored~\cite{darer2017filteredweb, lowe2007great,
levis2012collateral}. Each web page that is censored is fed back to
the beginning of the system. FilteredWeb discovered 1,355 censored
domains, 759 of which are outside the Alexa Top 1,000.

In this paper, we build upon the approach of FilteredWeb in the
following ways. First, {\em we extract content-rich phrases
for search queries}. In contrast, FilteredWeb only uses single words
for search queries. These phrases provide greater context regarding
the subject of censored web pages, which enables us to find websites
that are very closely related to each other. For example, consider the
phrase \begin{CJK*}{UTF8}{gbsn}中国侵犯人权 \end{CJK*} (Chinese human
rights violation). When we perform the searches \texttt{Chinese},
\texttt{human}, \texttt{rights} , and \texttt{violation}
independently, we mainly get websites for Western media outlets, many
of which are known to be censored in China. By contrast, if we search
for \texttt{Chinese human rights violation} as a single phrase, then
we discover a significant number of websites related to Chinese
culture, such as homepages for activist groups in China and
Taiwan. Identifying and extracting such key phrases is a non-trivial
task, as we discuss later.

Second, we use natural language processing to parse Chinese text
when adding to the blocklist. In contrast, FilteredWeb only extracts English
words that appear on a web page. As such, \textit{any website that is written
in simplified Chinese is ignored}, neglecting a significant portion of
censored sites. For example, there are many censored websites and blogs that
cover Chinese news and culture, and many of them only contain Chinese text. As
such, to discover region-specific, censored websites, such a system should be
able to parse Chinese text. Because Chinese is typically written without
spaces separating words, this requires the use of natural-language processing
tools.

Third, we make our block list public, in contrast to previous work. The
authors of FilteredWeb made their block list available to us for validation;
we have published our block list so others can build on
it~\cite{censorsearch-lists}.

In summary, we built and now maintain a large, public, culture-specific list
of websites that are censored in China. These websites cover topics such as
political dissent, historical events pertaining to the Chinese Communist
Party, Tibetan rights, religious freedom, and more. Furthermore, because many
of these website are written from the perspective of Chinese nationals and expatriates,
we are able to get first-hand accounts of Chinese culture that are
not present in other block lists. This new resource can help researchers who
are interested in studying Chinese censorship from the perspective of
marginalized groups that most affected by it.

In this paper, we make the following contributions:
\begin{itemize}
  \item We build upon the approach of FilteredWeb to discover censored
websites in China that specifically pertain to its culture. We do so
by extracting potentially sensitive Chinese phrases from censored
web pages and using them as search terms to find related websites.
  \item We build a list of 1125 censored domains in China, which is
12.5$\times$ larger than the standard list for censorship
measurements~\cite{citizenlab:block}. Furthermore, \textit{none of
these websites are on the largest block list
available}~\cite{darer2017filteredweb}.
  \item We perform a qualitative analysis of our block list to
    showcase its advantages over previous work.
\end{itemize}

The rest of the paper proceeds as follows. First, we describe our
approach to building a large, culture-specific block list for
China. This includes an in-depth analysis of the advantages of our
approach over previous work. Then, we describe three large-scale
evaluations that we performed. Each of these evaluations
produced qualitatively different results due to different
configurations of our system. Finally, we conclude
with a discussion of how the block list we built could be used by
researchers. We also briefly explore directions for future work.

\section{Related work}
Existing block lists have several limitations. First, some of these
block lists have been curated by organizations that study Internet
censorship, but these lists are now
outdated~\cite{chinadigitaltimes,oni}. Other lists have been
automatically generated by systems similar to ours, but they are not
publicly available at the time of
publication~\cite{sfakianakis2011censmon, darer2017filteredweb,
darer2018automated}. Furthermore, these systems do not focus on
blocked websites that are particular to Chinese culture. There are
also systems that create block lists through crowd-sourcing, but are
unable to automatically detect newly censored websites~\cite{herdict,
greatfire}. Finally, the Citizen Lab block list focuses on popular
websites--such as social media, Western media outlets, and VPN
providers--but not websites pertaining to Chinese
culture~\cite{citizenlab:block}.

Other systems use block lists to determine \textit{how} censorship
works, but they do not create more block lists. For example, Pearce
et al. proposed a system that uses block lists to measure how DNS
manipulation works on a global scale by combining DNS queries with AS
data, HTTPS certificates, and more~\cite{pearce2017global}. Pearce
et al. also built Augur, a system that utilizes TCP/IP side channels
between a host and a potentially censored website to determine whether
or not they can communicate with each
other~\cite{pearce2017augur}. Furthermore, Burnett et al. proposed
Encore, a system that utilizes cross-origin requests to potentially
censored websites to \textit{passively} measure how censorship works
from many vantage points~\cite{burnett:encore}. Lastly, several
platforms have been proposed that crowd-source censorship measurements
from users by having them install custom software on their
devices~\cite{razaghpanah2016exploring, ooni:about, iclab}.

\section{Approach}

\begin{figure}[t]
  \centering
  \includegraphics[scale=0.23]{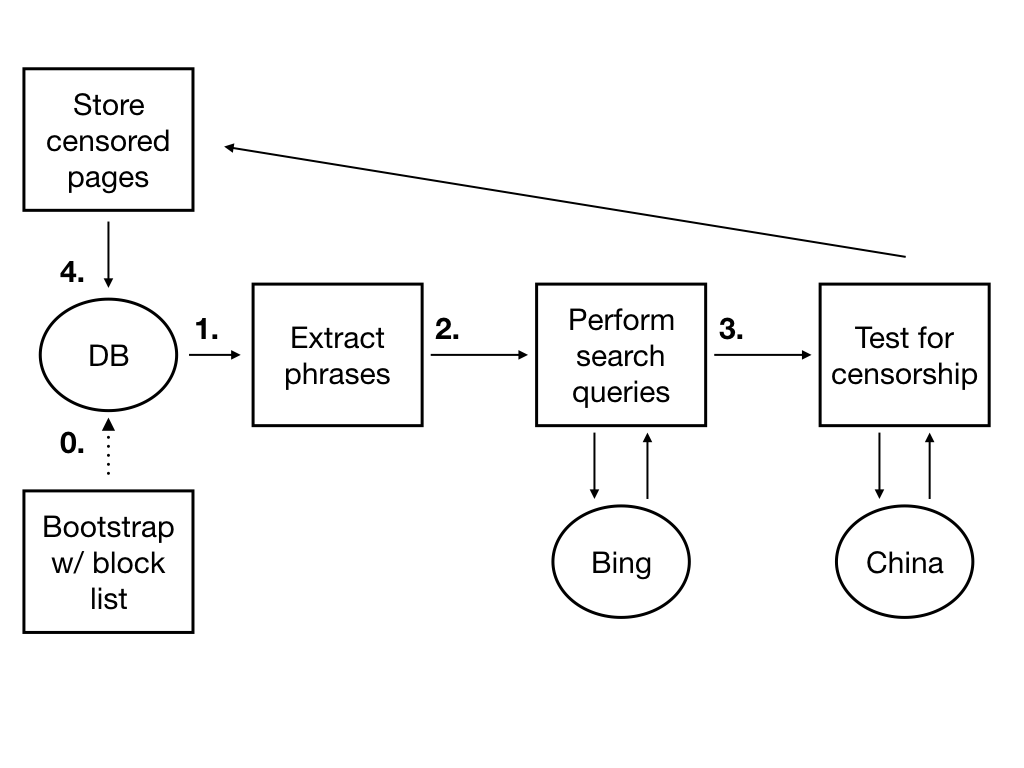}
  \caption{\label{arch}Our approach for discovering censored websites.}
\end{figure}

Figure~\ref{arch} summarizes our approach to finding censored
websites. For the most part, our approach is similar to that of
FilteredWeb~\cite{darer2017filteredweb}. We start by bootstrapping a
list of web pages that are known to be censored, such as the Citizen
Lab block list~\cite{citizenlab:block}.  Then, we extract Chinese and
English phrases that characterize these web pages.  Then, we use these
phrases as search queries to find related web pages that might also be
censored. Finally, we test each search result for DNS manipulation in
China and feed the censored web pages back to the beginning of the
system. The rest of this section details the new capabilities of our
approach beyond the state of the art.

\subsection{Extracting multi-word phrases}
An n-gram is a building block for natural-language processing that
represents a sequence of ``n'' units of text, e.g. words. For example,
if we were to compute all of the bigrams of words in the English
phrase \texttt{Chinese human rights violation}, we'd get
\texttt{Chinese human}, \texttt{human rights}, and \texttt{rights
violation}. Similarly, if we were to compute all the trigrams of
words, we'd get \texttt{Chinese human rights} and \texttt{human rights
violation}. On the other hand, if we simply extract unigrams from the
sentence, we'd be left with \texttt{Chinese}, \texttt{human},
\texttt{rights}, and \texttt{violation}. As such, we believe that by
extracting content-rich unigrams, bigrams, trigrams, we're able to
learn more about the subject of a web page.

Unfortunately, computing the n-grams of words in Chinese text is
difficult because it is typically written without spaces, and there is
no clear indication of where characters should be separated. Computing
such boundaries in natural-language processing tasks is often
probabilistic, and depending on where characters are separated, one
can arrive at very different meanings for a given
phrase~\cite{stanford:segmenter}. For example, consider the
text \begin{CJK*}{UTF8}{gbsn}特首\end{CJK*}. This is considered a
Chinese unigram that translates to ``chief executive'' in English, but
the character \begin{CJK*}{UTF8}{gbsn}特
\end{CJK*} on its own translates to ``special'', and the
character \begin{CJK*}{UTF8}{gbsn}首\end{CJK*} on its own translates
to ``first''~\cite{google:translate}. Given that we do not have
domain-specific knowledge of the web pages that we're analyzing, we
consider a unigram to be whatever the segmenter software in the
Stanford CoreNLP library considers to be a unigram, even if the output
consists of multiple English words.~\cite{tseng2005conditional,
chang2008optimizing}.

We believe that using Chinese unigrams, bigrams, and trigrams for search queries
instead of individual English words is more effective for discovering
censored websites in China. For example, we know that websites that
express collective dissent are considered sensitive by the Chinese
government~\cite{king2013censorship}. If we used the words
\texttt{destroy}, \texttt{the}, \texttt{Communist}, or \texttt{Party}
individually as search terms, we might not get websites that express
collective dissent because the words have been taken out of
context. On the other hand, the phrase ``\texttt{destroy the Communist
Party}'' as a whole expresses collective dissent, which might enable us
to find many censored domains when used as a search term. We evaluate
the effectiveness of such phrases in Section \ref{phrases-eval}.

\subsection{Ranking phrases} \label{tf-idf}
To ``rank'' the phrases on a censored web page, we use
term-frequency/inverse document-frequency (TF-IDF). TF-IDF is a
natural language technique that allows us to determine which phrases
best characterize a given web page~\cite{ramos2003using}. It can be
thought to work in three steps. First, we compute the term-frequency
for each phrase on a web page, which means that we count the frequency
of each unigram, bigram and trigram. Then, we compute the document-frequency
for each phrase on a web page, which entails searching a Chinese corpus
for the frequency of a given phrase across all documents in the
corpus~\cite{phrasefinder}. Finally, we multiply the term frequency by
the inverse of the document frequency. The resulting score gives us an
idea of how important a given Chinese phrase is to a web page.

Using this method, phrases like \begin{CJK*}{UTF8}{gbsn}1989年民主运
动\end{CJK*} [\texttt{1989 democracy movement}]
and \begin{CJK*}{UTF8}{gbsn}天安门广场示威\end{CJK*}
[\texttt{Tienanmen Square Demonstrations}] might rank highly on a
website about Chinese political protests. We would then use these phrases
as search terms in order to find related websites that might also be
censored. If we find a lot of censored URLs as a result, then we can
infer that the topic covered by that phrase is considered sensitive in China.

\subsection{Parsing Chinese text}
Before we can compute TF-IDF for a given web page, we need to tokenize the
text. Doing so is simple enough for English because each word is separated by a
space. For languages such as Chinese, however, all of the words in a sentence are
concatenated, without any spaces between them. As such, we need to apply
natural-language processing techniques to perform fine-grained analysis
of the text on a web page.

To do so, we make use of Stanford CoreNLP, a set of natural language
processing tools that operate on text for English, Arabic, Chinese, French,
German, and Spanish~\cite{corenlp2016suite}. For Chinese web pages, it allows us
split a sentence into a sequence of unigrams, each of which may represent
one word or even multiple words. By combining neighboring unigrams, then, we
can extract key phrases from a web page that describe its content.

As previously mentioned, although FilteredWeb is concerned with finding web pages
that are censored in China, it is only able to parse text for the ISO basic
Latin alphabet~\cite{darer2017filteredweb}. By being able to parse Chinese text as
well as ISO Latin text, then, we are able to cover many more web pages and
extract regional information that may explain why a given web page is
censored. In future work, we could make use of more intricate tools
from Stanford CoreNLP likes its part-of-speech tagger to identify
phrases that convey relevant information.

\begin{figure*}[t]
  \centering
  \includegraphics[scale=0.6]{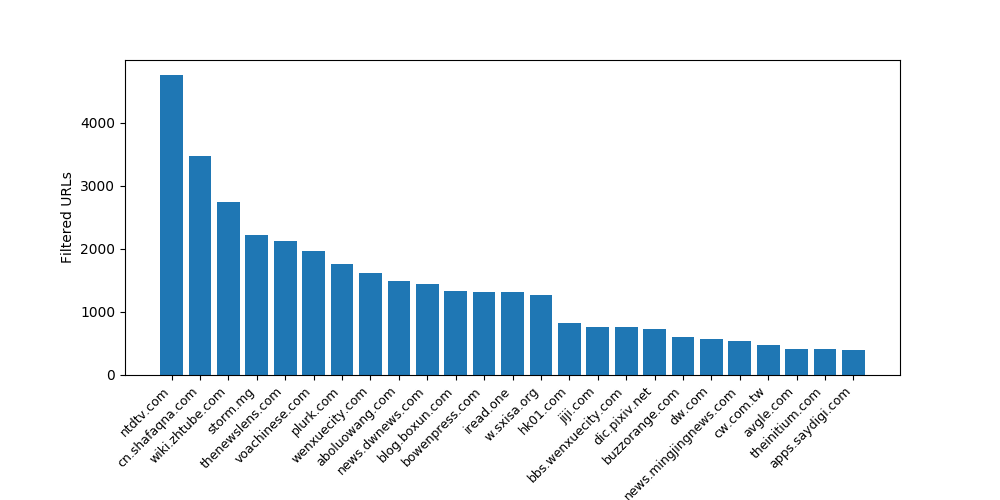}
  \caption{\label{top-domains}Top censored domains by URL count.}
\end{figure*}

\section{Evaluation}

We performed a large-scale evaluation of our approach to
discover region-specific websites that are censored in China.  We
began by seeding with the Citizen Lab block list, which is
the most widely used list by censorship researchers. The list contains
220 web pages that either are blocked or have been alleged to be blocked in
the past. Since we only extract phrases from web pages that are
currently censored, we began by testing each web page on the block list for
censorship. This left us with 108 unique web pages and 85 domains.

From November 11th, 2017 to July 9th, 2018, we used Bing's Search API
to search for websites related to known censored
websites~\cite{microsoft:bing}. According to the API, each call can
return at most 50 search results. Since it would be expensive to
perform multiple API calls for each search term, we limited ourselves
to one API call, i.e. 50 search results, per search term. For each
website in the search results, we tested for censorship by sending a
DNS request to a set of controlled IP addresses in China that don't
belong to DNS servers. As with FilteredWeb, if we received a DNS
response, we inferred that the DNS request was intercepted, and thus
the tested website is censored. Finally, we performed three separate
evaluations to measure the effectiveness of different phrase sizes for
finding censored websites. With each evaluation, we only extracted
unigrams, bigrams, and trigrams, respectively. We also limited each
evaluation to 1,000,000 URLs to be consistent with the methodology of
FilteredWeb~\cite{darer2017filteredweb}.

We also configured the Bing API calls so that any URLs from Blogspot,
Facebook, Twitter, YouTube, and Tumblr would be ignored. We did this
for a couple of reasons. For one, these websites are widely known to
be censored in China, so we would not be providing new information by
having these websites or their subdomains in our result
set. Furthermore, Tumblr and Blogspot assign a unique subdomain to
each user's blog, and in some cases we'd get a dozen blogs from a
single search query. In order to find culture-specific websites that
are normally buried by the top 50 search results, then, we need to
omit these blogs.
\begin{figure*}[t]
  \centering
  \includegraphics[scale=0.6]{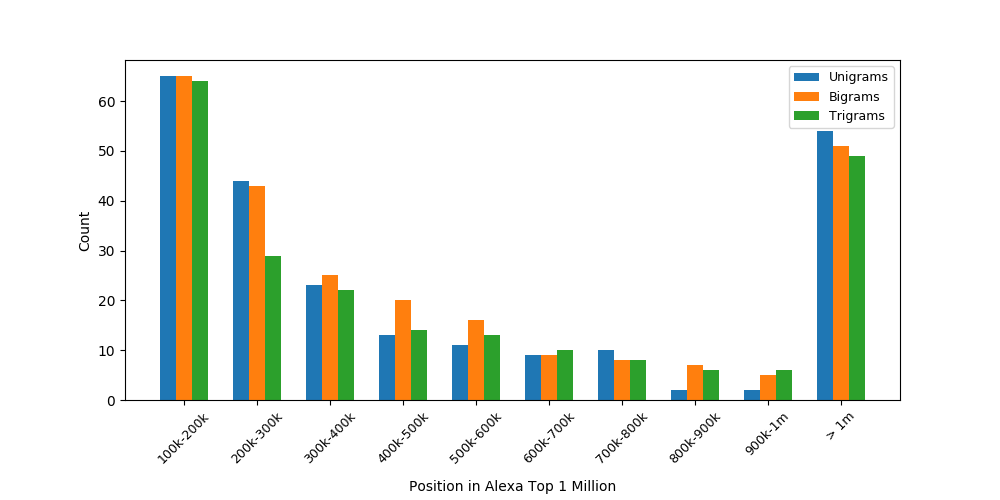}
  \caption{\label{alexa}Ranking of censored websites on Alexa Top 1 Million.}
\end{figure*}

\section{Results}
We reached several key insights from performing our
evaluations. First, we were able to discover hundreds of
censored websites that are not present on existing block
lists. Furthermore, we noticed that many websites on our block list
receive very little traffic. Lastly, we found that by using
politically-charged phrases as search terms, we were able to find a
disproportionately large amount of censored websites. The rest of this
section discusses the main findings in depth.

\subsection{Existing blocklists are incomplete}
\textit{By using natural-language processing on Chinese web pages, we
were able to discover hundreds of censored websites that are not
present on the Citizen Lab block list--the standard for censorship
measurements--and FilteredWeb's block list~\cite{darer2017filteredweb,
citizenlab:block}}. Furthermore, only 3 of the top 50 censored domains
that appeared in our search results are in the top 50 censored domains
found by FilteredWeb. Figure~\ref{top-domains} breaks down the URL
count for the top 25 censored domains that we discovered. These
websites seem to mainly cover Chinese human rights issues, news,
censorship circumvention, and more.

For instance, \texttt{wiki.zhtube.com} appears to be a Chinese
Wikipedia mirror. The web page for \texttt{zhtube.com} seems to be the
default page for a Chinese LNMP (Linux, NginX MySQL, PhP) installation
on a virtual private server~\cite{lnmp}. We found over 3000 URLs in
our dataset that point to this website, which may suggest that Chinese
websites are trying to circumvent the on-and-off censorship of
Wikipedia in China~\cite{wikipedia-china}. Furthermore,
\texttt{blog.boxun.com} is a United States-based outlet for Chinese
news that relies heavily on anonymous submissions. The owners of the
website note that ``Boxun often reports news that authorities do not
tell the public, such as outbreaks of diseases, human rights
violations, corruption scandals and
disasters''~\cite{boxun-about}. \textit{Thus, by building on the
approach of FilteredWeb, we were able to produce a qualitatively
different block list}. We recommend putting these two block lists
together to create a single block list that is both wide in scope and
large in size.

Intuitively, we were able to find hundreds of qualitatively
different censored domains than those found by FilteredWeb because
we could extract Chinese-specific topics from web pages. We were able
to do so by using TF-IDF with a Chinese corpus to rank the
``uniqueness'' of Chinese phrases that appeared on a given
web page. For example, the trigram
\begin{CJK*}{UTF8}{gbsn}仅 限于 书面\end{CJK*} (``only written'') was poorly ranked on
\texttt{tiananmenmother.org}--a Chinese democratic activist group--
because although it appears frequently, it's a common phrase in
Chinese, according to the Chinese corpus on
PhraseFinder~\cite{phrasefinder}. On the other hand, the
trigram \begin{CJK*}{UTF8}{gbsn}自由 亚洲 电台\end{CJK*} (``Radio Free
Asia'') was highly ranked because it didn't appear frequently, and
it's an uncommon phrase in Chinese. It should also be noted that Radio
Free Asia is a broadcasting corporation whose stated mission is to
``provide accurate and timely news and information to Asian countries
whose governments prohibit access to a free
press''~\cite{rfa:about}. Thus, by scoring Chinese phrases with TF-IDF
against a Chinese corpus, we were able to determine which phrases are
``content-rich'' on a given web page.

\begin{table}[b]
  \begin{center}
    \scalebox{0.80}{
    \begin{tabular}{ | l | r | r | r | }
      \hline
      Phrase length & All censored domains & New censored domains \\ \hline
      Unigrams      & 1029    & 719 \\
      Bigrams       & 970     & 598 \\
      Trigrams      & 975     & 579 \\
      Total         & 1756    & \textbf{1125} \\
      \hline
    \end{tabular}}
  \end{center}
\caption{\label{breakdown} Total number of censored domains discovered.}
\end{table}

\subsection{China blocks many unpopular websites}
Figure~\ref{alexa} shows the ranking of the websites we discovered on
the Alexa Top 1,000,000. Notably, many of the websites we discovered
are spread throughout the tail of the list, and some of the websites
are not even on the list at all. \textit{Given that the top 100,000 websites
likely receive the vast majority of traffic on the Internet, we can
infer that censors in China are not just interested in blocking
``big-name'', popular websites.} They are actively seeking out websites
of \textit{any size} that contain ``sensitive'' content. We also
discovered a number of websites that fall outside the Alexa Top
1,000,000 altogether. Without the use of an automated system that can
discover censored websites, it's unlikely that the public would even
be aware that these websites are blocked.

\begin{figure}[t]
  \centering
  \includegraphics[scale=0.5]{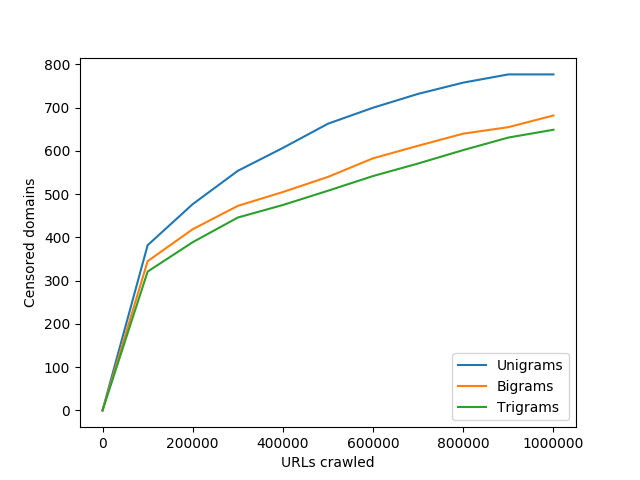}
  \caption{\label{censored-vs-urls} Censored domains discovered over unique URLs crawled. }
\end{figure}

\begin{table}[b]
  \begin{center}
    \scalebox{0.9}{
      \begin{tabular}{ l | l | c }
        Chinese & English & Censored domains \\ \hline
        \begin{CJK*}{UTF8}{gbsn}王歧山\end{CJK*} & Wang Qishan & 74\% \\
        \begin{CJK*}{UTF8}{gbsn}李洪志\end{CJK*} & Li Hongzhi & 64\% \\
        \begin{CJK*}{UTF8}{gbsn}郭伯雄\end{CJK*} & Guo Boxiong & 62\% \\
        \begin{CJK*}{UTF8}{gbsn}胡锦涛\end{CJK*} & Hu Jintao & 56\% \\
        \begin{CJK*}{UTF8}{gbsn}胡平\end{CJK*}  & Hu Ping & 54\% \\
        \begin{CJK*}{UTF8}{gbsn}Morty\end{CJK*} & Morty & 52\% \\
        \begin{CJK*}{UTF8}{gbsn}命案\end{CJK*}  & Homicide & 52\% \\
        \begin{CJK*}{UTF8}{gbsn}特首\end{CJK*}  & Chief executive & 52\% \\
        \begin{CJK*}{UTF8}{gbsn}Vimeo\end{CJK*} & Vimeo & 50\% \\
        \begin{CJK*}{UTF8}{gbsn}中情局\end{CJK*} & CIA & 50\% \\
      \end{tabular}}
  \end{center}
  \caption{\label{effective-unigrams}Sample of unigrams with
    significant blockrates}
\end{table}

Table \ref{breakdown} shows the breakdown of how many censored
websites we discovered. By using unigrams, we were able to discover
719 domains that are not on any publicly available block list. By
using bigrams, we were able to discover 598 censored domains. Lastly,
by using trigrams, we were able to discover 579 censored domains. In
total, we discovered 1125 censored domains, none of which are on the
Alexa Top 1000 or Darer et al.'s blocklist~\cite{darer2017filteredweb,
  alexa:top1000}. Each evaluation found 273
censored domains in common. Furthermore, each of these evaluations
were performed with 1,000,000 unique URLs, consistent with the
methodology of FilteredWeb~\cite{darer2017filteredweb}. Figure
\ref{censored-vs-urls} shows how many censored domains we discovered
as a function of unique URLs crawled for each evaluation.

\begin{table*}[t]
  \begin{center}
    \scalebox{0.9}{
      \begin{tabular}{ l | l | c }
        Chinese & English & Censored domains \\ \hline
        \begin{CJK*}{UTF8}{gbsn}中共\end{CJK*} \begin{CJK*}{UTF8}{gbsn}威胁\end{CJK*} & Chinese Communists threaten & 50\% \\
        \begin{CJK*}{UTF8}{gbsn}声明的\end{CJK*} \begin{CJK*}{UTF8}{gbsn}反共产主义\end{CJK*} & Declared anti-communist & 44\% \\
        \begin{CJK*}{UTF8}{gbsn}中国共\end{CJK*} \begin{CJK*}{UTF8}{gbsn}产党的公共安全\end{CJK*} & Public security of the CPC & 42\% \\
        \begin{CJK*}{UTF8}{gbsn}北京\end{CJK*} \begin{CJK*}{UTF8}{gbsn}清洁\end{CJK*} & Beijing clean-up & 40\% \\
        \begin{CJK*}{UTF8}{gbsn}江泽民\end{CJK*} \begin{CJK*}{UTF8}{gbsn}胡锦涛\end{CJK*} & Jiang Zemin Hu Jintao & 40\% \\
        \begin{CJK*}{UTF8}{gbsn}迫害\end{CJK*} \begin{CJK*}{UTF8}{gbsn}活动\end{CJK*} & Persecution
                                                     activities & 40\% \\
        \begin{CJK*}{UTF8}{gbsn}官员\end{CJK*} \begin{CJK*}{UTF8}{gbsn}呼吁\end{CJK*} & Officials called on & 36\% \\
        \begin{CJK*}{UTF8}{gbsn}重的\end{CJK*} \begin{CJK*}{UTF8}{gbsn}公民\end{CJK*} & Heavy Citizen & 36\% \\
        \begin{CJK*}{UTF8}{gbsn}非法\end{CJK*} \begin{CJK*}{UTF8}{gbsn}拘留\end{CJK*} & Illegal detention
                          & 36\% \\
        \begin{CJK*}{UTF8}{gbsn}不同\end{CJK*} \begin{CJK*}{UTF8}{gbsn}的民主\end{CJK*} & Different Democratic & 34\% \\
      \end{tabular}}
  \end{center}
  \caption{\label{effective-bigrams}Sample of bigrams with significant
    block rates}
\end{table*}

\begin{table*}[t]
  \begin{center}
    \scalebox{0.9}{
      \begin{tabular}{ l | l | c }
        Chinese & English & Censored domains \\ \hline
        \begin{CJK*}{UTF8}{gbsn}北戴\end{CJK*} \begin{CJK*}{UTF8}{gbsn}河\end{CJK*} \begin{CJK*}{UTF8}{gbsn}会议\end{CJK*} & BEIDAIHE meeting & 54\% \\
        \begin{CJK*}{UTF8}{gbsn}中国\end{CJK*} \begin{CJK*}{UTF8}{gbsn}共产党\end{CJK*} \begin{CJK*}{UTF8}{gbsn}的宗教政策\end{CJK*} & The Chinese Communist Party's religious policy & 42\% \\
        \begin{CJK*}{UTF8}{gbsn}采取\end{CJK*} \begin{CJK*}{UTF8}{gbsn}暴力\end{CJK*} \begin{CJK*}{UTF8}{gbsn}镇压\end{CJK*} & To take a violent crackdown & 38\% \\
        \begin{CJK*}{UTF8}{gbsn}香港\end{CJK*} \begin{CJK*}{UTF8}{gbsn}政\end{CJK*} \begin{CJK*}{UTF8}{gbsn}治\end{CJK*} & Hong Kong Politics & 34\% \\
        \begin{CJK*}{UTF8}{gbsn}欧洲\end{CJK*} \begin{CJK*}{UTF8}{gbsn}议会\end{CJK*} \begin{CJK*}{UTF8}{gbsn}决议\end{CJK*} & European Parliament Resolution & 32\% \\
        \begin{CJK*}{UTF8}{gbsn}新\end{CJK*} \begin{CJK*}{UTF8}{gbsn}唐\end{CJK*} \begin{CJK*}{UTF8}{gbsn}王朝\end{CJK*} & New Tang Dynasty & 32\% \\
        \begin{CJK*}{UTF8}{gbsn}恐怖\end{CJK*} \begin{CJK*}{UTF8}{gbsn}事件\end{CJK*} \begin{CJK*}{UTF8}{gbsn}。\end{CJK*} & A terrorist event. & 32\% \\
        \begin{CJK*}{UTF8}{gbsn}天安\end{CJK*} \begin{CJK*}{UTF8}{gbsn}门广\end{CJK*} \begin{CJK*}{UTF8}{gbsn}场示威\end{CJK*} & Tienanmen Square Demonstrations & 32\% \\
        \begin{CJK*}{UTF8}{gbsn}敦促\end{CJK*} \begin{CJK*}{UTF8}{gbsn}美国\end{CJK*} \begin{CJK*}{UTF8}{gbsn}政府\end{CJK*} & Urging the US government & 32\% \\
        \begin{CJK*}{UTF8}{gbsn}1989\end{CJK*} \begin{CJK*}{UTF8}{gbsn}民主\end{CJK*} \begin{CJK*}{UTF8}{gbsn}运动\end{CJK*} & 1989 democracy movement & 30\% \\
      \end{tabular}}
  \end{center}
  \caption{\label{effective-trigrams}Sample of trigrams with
    significant blockrates}
\end{table*}

\subsection{Political phrases are highly effective}
\label{phrases-eval}

We also wanted to see if there is a correlation between the
presence of certain phrases and whether or not a given website is
censored, even if we have no ground truth.  For example, if we make a search
for \begin{CJK*}{UTF8}{gbsn}中国侵犯人权\end{CJK*} (Chinese human
rights violation) and find that a particular search result is
censored, then we cannot be certain whether the presence of that phrase
\textit{caused} the website to be blocked. Even if we assume 
that a censor is manually combing through search engines to find
sensitive websites, the website could have been blocked because it
contained totally different content.

Nevertheless, there seems to be some correlation for certain
phrases. Table~\ref{effective-unigrams} shows a sample of unigrams
that returned the most number of unique filtered domains from
Bing. For example, the top four unigrams--Wang Qishan, Li Hongzhi, Guo
Boxiong, and Hu Jintao--are the names of controversial figures in
Chinese history. Li Hongzhi is the leader of the Falun Gong spiritual
movement, whose practitioners have been subject to persecution and
censorship by the Chinese government since
1999~\cite{freedomhouse:falun}. Similarly, Guo Boxiong is a former top
official of the Chinese military that was sentenced to life in
prison in 2016 for accepting bribes, according to the Chinese
government~\cite{guardian:guo}. Thus, if a Chinese website discusses
these people, the website may become censored for containing
``sensitive'' content.

There are also bigrams that correlate with a large percentage of
censored domains, but they are more explicitly political than the
unigrams. Table~\ref{effective-bigrams} shows this result. First, most
of the bigrams refer to the Chinese Communist Party in some
way. Phrases such as \begin{CJK*}{UTF8}{gbsn}中共的威胁\end{CJK*}
(Chinese Communists threaten), \begin{CJK*}{UTF8}{gbsn}江泽民胡锦
涛\end{CJK*} (Jiang Zemin Hu Jintao), and \begin{CJK*}{UTF8}{gbsn}中国
共产党的治安\end{CJK*} (Public security of the CPC) do not necessarily
convey sensitive information, but they nonetheless refer to the
government of China. On the other hand, some phrases clearly refer to
political dissent, such as \begin{CJK*}{UTF8}{gbsn}官员呼吁\end{CJK*}
(Officials called on), \begin{CJK*}{UTF8}{gbsn}迫害活动\end{CJK*}
(Persecution activities), \begin{CJK*}{UTF8}{gbsn}非法拘留\end{CJK*}
(Illegal detention), and \begin{CJK*}{UTF8}{gbsn}宣称反共\end{CJK*}
(Declared anti-communist).

We see a similar result with trigrams, as shown by
Table~\ref{effective-trigrams}. Phrases that stand out
include \begin{CJK*}{UTF8}{gbsn}中国共产党的宗教政策\end{CJK*} (The
Chinese Communist Party's religious policy), \begin{CJK*}{UTF8}{gbsn}
天安门广场示威\end{CJK*} (Tienanmen Square
demonstrations), \begin{CJK*}{UTF8}{gbsn}1989年民主运动\end{CJK*}
(1989 democracy movement), and \begin{CJK*}{UTF8}{gbsn}采取暴力镇
压\end{CJK*} (To take a violent crackdown). Interestingly, we also see
that discussion of China's religious policy, the ``New Tang
Dynasty''--a religious radio broadcast located in the United States--,
and European Union legislation may also be considered sensitive
content.~\cite{china-religion}. \textit{Together, these results
suggest that references to collective political dissent are highly
likely to be censored}. This is consistent with the findings of King
et al.~\cite{king2013censorship}.

\balance
\section{Conclusion}
We built a block list of 1125 censored websites in China that are not
present on the largest block
list~\cite{darer2017filteredweb}. Furthermore, our list is
12.5$\times$ larger than the most widely used block
list~\cite{citizenlab:block}. It contains human rights organizations,
minority news outlets, religious blogs, political dissent groups,
privacy-enhancing technology providers, and more. We've made our
source code and block list available on
GitHub~\cite{censorsearch-lists}.

Automatically detecting which websites are censored in a given country
is an open problem.  One way of improving our work would be to
experiment with advanced natural-language processing techniques to
identify better search terms. For example, we could try using Stanford
CoreNLP's part-of-speech tagger to identify phrases that describe some
action against the Chinese government. This approach might identify
the following phrase: ``Chinese citizens protest against the Communist
Party on June 4th''. We believe that using such culture-specific
phrases as search terms would enable us to discover even more websites
that are censored in China.

\section*{Acknowledgments}
We thank the anonymous FOCI '18 reviewers for their helpful
feedback. This research was supported by NSF awards CNS-1409635,
CNS-1602399, and CNS-1704105.

\newpage

{\footnotesize
 \bibliographystyle{acm}
 \balance\bibliography{../common/bibliography}}

\end{document}